\documentclass[reprint,twocolumn,superscriptaddress,showpacs,nofootinbib,notitlepage,preprintnumbers]{revtex4-2}

\usepackage{graphicx}
\usepackage{hyperref}
\usepackage[dvipsnames]{xcolor}
\usepackage{amsmath,amsthm,amssymb}
\usepackage{comment}
\usepackage{enumerate}
\newcommand{\todo}{\colorbox{pink}{\textsc{Todo}}}

\DeclareMathOperator{\Tr}{Tr}
\DeclareMathOperator{\tr}{tr}

\newtheorem*{theorem*}{Theorem}

\newtheorem*{problem*}{Problem}

\usepackage[all,vshift=-1em,color=Black,angle=90,scale=3]{background}
\SetBgContents{\textsc{Draft}}
\SetBgPosition{current page.west}
\SetBgContents{}

\begin{document}
\preprint{LA-UR-24-23391}

\title{Schwinger-Dyson control variates for lattice fermions}

\author{Scott Lawrence}
\email{srlawrence@lanl.gov}
\affiliation{Los Alamos National Laboratory Theoretical Division T-2, Los Alamos, NM 87545, USA}

\date{\today}

\begin{abstract}
	Previous work has shown that high-quality control variates for lattice Monte Carlo methods may be constructed from lattice Schwinger-Dyson relations. This paper extends that method to theories with lattice fermions, using the Thirring model in $1+1$ spacetime dimensions as a testbed. Past construction of these control variates involved a number of fitting parameters that scaled with lattice volume. By computing the control variate in perturbation theory, the number of fitting parameters required for an order-of-magnitude improvement in the signal-to-noise ratio at weak coupling is reduced to be of order one.
\end{abstract}

\maketitle

\section{Introduction}\label{sec:introduction}
Lattice Monte Carlo is the predominant method for first-principles calculations in strongly coupled field theories. When applied to theories and regimes in which the Euclidean action is real---including QCD at vanishing chemical potential---a fixed observable may in practice be measured in time polynomial in the size of the lattice being simulated and the desired precision. The polynomial coefficients may however be large, and even parametrically large in properties of the observable.

As is typical for Monte Carlo methods, the number of samples required to measure an expectation $\langle \mathcal O \rangle$ to precision $\epsilon$ is proportional to $\epsilon^{-2}$. The coefficient of proportionality is the standard deviation of the observable being measured, $\sigma = \sqrt{\langle \mathcal O^2 \rangle - \langle \mathcal O\rangle^2}$. This scaling become problematic in cases where the noise of the observable is large. A classic example is in the baryon correlation function in lattice QCD, where the signal-to-noise ratio falls exponentially with separation~\cite{Parisi:1983ae,Lepage:1989hd}.

The other famous obstacle to calculations in lattice field theory, namely the sign problem, has a naive difficulty which scales exponentially with the spacetime volume being simulated. The number of samples required scales proportionally to $e^{2 (m L)^d}$, with $d$ the lattice dimension and $m$ a mass scale characteristic of the problem. In four dimensions, it is clear that a calculation on even a modestly sized lattice will suffer catastrophically. In the case of the Thirring model in three dimensions, a $10^3$ lattice in the heavy-dense limit would require $\sim 10^{380}$ samples to resolve the average phase~\cite{Lawrence:2018mve}.

In contrast, many typical signal-to-noise problems scale proportional to a length, rather than a spacetime volume. Consequently these are far more accessible. A speedup of a factor of $10$ is irrelevant for sign problems of physical interest, but practically useful for improving the signal-to-noise ratio of a correlator at somewhat longer time separations. The purpose of this paper is to show how such improvements in the signal-to-noise ratio may be obtained in theories with lattice fermions, through the construction of suitable \emph{control variates}.

A control variate is a function $f$ of the field configurations whose expectation value is guaranteed to vanish, but which is sample-by-sample correlated with an observable $\mathcal O$ of physical interest. The modified observable $\tilde{\mathcal O} \equiv \mathcal O - f$ therefore has the same expectation value as the original, but reduced variance. As a result, once a suitable control variate has been identified, such a modified observable may be measured in lieu of $\mathcal O$, yielding a higher signal-to-noise ratio.

The idea of using control variates to accelerate lattice Monte Carlo calculations is not a new one. In contexts where sampling is cheap relative to computation of the observable, the approximant method was proposed in which a cheap approximation to the observable is evaluated on many samples, while the difference between that observable and the true one is computed on only a few~\citep{Blum:2012uh,Yoon:2018krb}. The method discussed here is applicable instead to contexts where sampling itself is the primary bottleneck, and therefore no observable can be computed cheaply with high statistics. Control variates of this sort for lattice systems have been constructed in perturbation theory~\cite{Lawrence:2020kyw} and via machine learning~\cite{Lawrence:2022dba,Lawrence:2023cft,Bedaque:2023ovz}

This work follows from the observation made in~\cite{Bhattacharya:2023pxx} that control variates can be constructed systematically via lattice Schwinger-Dyson relations. In that case, a high-dimensional space of Schwinger-Dyson relations was identified, and a high-dimensional fit performed to determine the most effective control variate. The dimension of the fit scaled linearly (potentially worse) with the volume of the system, requiring at least as many samples as lattice degrees of freedom. In the context of lattice QCD this is not a feasible approach. In this paper we will see that control variates can be constructed in perturbation theory with no fit needed. Moreover, by using the functional form obtained from perturbation theory, a few-parameter fit can be performed to yield control variates as effective as those obtained via many-parameter optimization.

Throughout this paper we will use the lattice Thirring model in $1+1$ spacetime dimensions, with staggered fermions, as our testbed. This model is detailed in Section~\ref{sec:thirring}. Lattice Schwinger-Dyson relations, for both bosonic and fermionic systems, are detailed in Section~\ref{sec:sd}, and used to construct general control variates for the Thirring model. In this framework, perturbatively optimal control variates are constructed and tested in Section~\ref{sec:perturbative}. Section~\ref{sec:fitting} demonstrates non-perturbative fits to forms of control variates inspired by perturbation theory. Section~\ref{sec:discussion} concludes with some further discussion.

\section{Thirring model}\label{sec:thirring}
Our testbed throughout this work will be a lattice discretization of the Thirring model in $1+1$ dimensions, with staggered fermions. On a lattice with $V$ sites, there are $2V$ degrees of freedom, labelled $A_\mu(x)$ for $\mu \in \{1,2\}$. This field may be thought of as being valued on $[0,2\pi)$ with the action and all observables periodic; alternatively it is the exponent of a $U(1)$-valued field. This latter phrasing makes clear the connection between the Thirring model and QED: this lattice Thirring model is a variant of compact QED without gauge invariance, and with an infinitely massive photon.

After fermions have been integrated out, the lattice action is
\begin{equation}\label{eq:thirring-action}
	S(A) = -\frac{2}{g^2} \sum_{x,\mu} \cos A_\mu(x) - \log \det D(A)
\end{equation}
where the Dirac matrix is defined as
\begin{widetext}
	\begin{equation}\label{eq:dirac}
	D_{y,x} =
	m \delta_{x,y}
	- \delta_{x,y+\hat 2} \frac{e^{-i A_2(y)}}{2} (-1)^{\delta_{x_2,0}}
	+ \delta_{x+\hat 2,y} \frac{e^{i A_2(x)}}{2} (-1)^{\delta_{y_2,0}}
	- \delta_{x,y+\hat 1} \frac{e^{-i A_1(y)}}{2}\eta(x)
	+\delta_{x+\hat 1,y} \frac{e^{i A_1(x)}}{2}\eta(x)
	\text.
\end{equation}
\end{widetext}
The lattice model is characterized by the bare mass $m$ and the bare coupling $g^2$. The staggered fermions are defined by the factor $\eta(x) = (-1)^{x_2}$. The staggered fermions reduce the number of doublers, so that the model above has two flavors of fundamental fermion.

The low-lying excitations are a fermion (a dressed version of the bare field $\psi)$ and a pseudoscalar bound state typically termed a `meson' by analogy with QCD. At weak coupling the meson mass $m_B$ is twice that of the fermion mass $m_F$. In the case of the one-flavor Thirring model, the strong coupling limit is dual to the weak coupling limit of the Sine-Gordon model~\cite{Coleman:1974bu}. For this reason, even for the case $N_f > 1$, the strong-coupling regime of the Thirring model often interpreted to be where $m_F \sim m_b$.

For concreteness, we will work primarily with only two sets of bare parameters: one corresponding to weak coupling in the continuum, and the other to strong coupling. The weak coupling parameters are $m=0.25$ and $g^2 = 0.1$, resulting in a fermion mass (in lattice units) $m_f \approx 0.27(1)$ and a boson mass approximately twice that. The strong coupling parameters are $m=0.05$ and $g^2 = 0.8$, yielding a fermion mass $m_f \approx 0.25(2)$ and a boson mass $m_b \approx 0.30(3)$; this second parameter set sits firmly in the strong-coupling regime.

\section{Schwinger-Dyson on the lattice}\label{sec:sd}
Our task, as described in the introduction, is to find a function $f$ for which, first, $\langle f \rangle = 0$, and second, the noise in $\tilde{\mathcal O} = \mathcal O - f$ is much reduced from that in $\mathcal O$.

We beginning by constructing a broad class of functions $f$ with vanishing expectation value.
Given a lattice action $S(\phi)$ of some bosonic fields $\phi$, an exact relation may be derived from any suitable function of the fields $g(\phi)$. First note that, with mild conditions on $g(\cdot)$ and the integration domain $\Phi$, the integral of a total derivative vanishes:
\begin{equation}
	0 = \int_\Phi d\phi \,\partial \left(e^{-S(\phi)} g(\phi)\right)
	\text.
\end{equation}
As a direct consequence we obtain the equality of two expectation values:
\begin{equation}\label{eq:sd}
	\langle \partial g \rangle = \langle g \partial S \rangle
\end{equation}
These are the \emph{lattice Schwinger-Dyson relations}, immitating the terminology of~\cite{Peskin:1995ev} in the continuum case. Note that this is not just one relation, but a linear space of relations with dimension equal to the dimension of the integration domain $\Phi$: Eq.~(\ref{eq:sd}) is true for any differentiation operator.

The condition on $g(\cdot)$ is only that it decay sufficiently quickly near the boundary $\partial \Phi$ of the domain of integration that boundary terms vanish when integrating by parts. In the context of scalar field theory, the Boltzmann factor decays exponentially, and it is therefore sufficient for $g(\phi)$ to be a polynomial. Where the domain of integration has no boundary (as is the case for gauge theories and the Thirring model), we require only that $g$ be differentiable.

\subsection{Virial theorem}

One instance of a Schwinger-Dyson control variate is already well known in another form: the use of the virial theorem to construct an alternative estimator of the kinetic energy.

A paradigmatic example is as follows. Consider a lattice Monte Carlo simulation of a single particle in a one-dimensional anharmonic trap. The (Euclidean) lattice action is
\begin{equation}
	S_{\mathrm{osc}}(x) = \sum_t \frac {(x_t - x_{t+1})^2}{2 a} + a V(x_t)
	\text.
\end{equation}
The lattice spacing is $a$. The details of the potential $V(x)$ will not be relevant. We want to estimate the kinetic energy; that is, the expectation value $\frac 1 2 \langle p^2 \rangle$. The corresponding function of lattice fields is found by evaluating $\langle x'|e^{-\frac a 2 p^2}p^2| x \rangle$, and is
\begin{equation}
	K_t[x] = \frac {1}{2a} - \frac{(x_{t+1} - x_t)^2}{2 a^2}
	\text.
\end{equation}
Unfortunately this observable is quite noisy, and moreso as $a \rightarrow 0$. Instead, one can evaluate~\cite{bartholomew1984fluctuations}
\begin{equation}
	K^{\text{(virial)}}_t[x] = \frac 1 2 \langle x_t V'(x_t) \rangle
	\text,
\end{equation}
which has, by the virial theorem, the same expectation value, but greatly reduced noise.

The equivalence of these two estimators can be shown directly, via a control variate constructed as described above. The relevant Schwinger-Dyson identity is
\begin{equation}
	\Big\langle \frac{\partial}{\partial x_t} \frac{x_t}{2a} \Big\rangle
	=
	\Big\langle \frac{x_t}{2a} \frac{\partial S}{\partial x_t} \Big\rangle
	\text.
\end{equation}
This holds for any time $t$. Expanding, we find that
\begin{equation}
	\frac 1 {2a}
	=
	\Big\langle
	\frac{x_t}{2a^2} \big(2x_t - x_{t+1} - x_{t-1}\big)
	\Big\rangle
	+
	\frac 1 2 \langle x_t V'(x_t)\rangle
	\text.
\end{equation}
Finally averaging over all translations of the lattice, we obtain the equivalence of $K_t$ and $K_t^{\text{(virial)}}$ as desired.

\subsection{Fermionic models}\label{ssec:fermionic-sd}

Such lattice Schwinger-Dyson relations may also be obtained from the Thirring model. The most straightforward approach is to begin with the action Eq.~(\ref{eq:thirring-action}) in which the fermions have been integrated out. This action is a function exclusively of the bosonic vector field $A$. Given a function $g(A)$ we have
\begin{equation}
	\Big\langle \frac{\partial}{\partial A_\mu(x)} g(A) \Big\rangle = \Big\langle g(A) \frac{\partial}{\partial A_\mu(x)} S(A) \Big\rangle
\end{equation}
for every site $x$ and direction $\mu$. Because the domain of integration is compact, there is no constraint (beyond differentiability) on the functions $g(A)$ that generate Schwinger-Dyson relations.

It will be helpful throughout what follows to express a Schwinger-Dyson relation as a function $f(\cdot)$ whose expectation value is guaranteed to vanish. Schwinger-Dyson relations of the form of Eq.~(\ref{eq:sd}) result in such functions 
\begin{equation}\label{eq:cv-general}
	f = \partial g - g \partial S\text.
\end{equation}
In the case above, that function with vanishing expectation value is
\begin{widetext}
\begin{equation}\label{eq:cv-A}
	F_{\mu}^{(x)}[g](A) = \frac{\partial g(A)}{\partial A_\mu(x)}
	- g(A) \frac{2}{g^2} \sin A_\mu(x) - g(A) \tr D^{-1} \frac{\partial D(A)}{\partial A_\mu(x)}
	\text.
\end{equation}
\end{widetext}

The action of Eq.~(\ref{eq:thirring-action}) is obtained from a more physical action with a four-fermi interaction term and no bosonic field. We can attempt to derive Schwinger-Dyson relations at any stage in this derivation. To begin with, consider the form of the path integral immediately before fermions have been integrated out. The action is
\begin{equation}
	S(A,\bar\psi,\psi) = -\frac 2 {g^2}\sum_{x,\mu}\cos A_\mu(x) - \bar \psi D(A) \psi
	\text.
\end{equation}
At this stage there are two ways in which we can attempt to derive Schwinger-Dyson relations. Entirely analogously to what is above, we can write
\begin{equation}
	0 = \int dA d\bar\psi d\psi\, \frac{\partial}{\partial A_\mu(x)}\left[ g(A) e^{-S(A,\bar\psi,\psi)}\right]\text.
\end{equation}
The resulting Schwinger-Dyson equations directly involve the Grassmann fields $\bar\psi$,$\psi$, and are not usable in a lattice calculation until fermions have been again integrated out. Upon integrating the fermions out, precisely those relations described by Eq.~(\ref{eq:cv-A}) are recovered.

Alternatively, we might integrate by parts using the Grassmann variables. Just as with ordinary integration (although for different reasons), Grassmann integrals obey the identity
\begin{equation}
	\int d\psi \frac{\partial}{\partial \psi} f(\psi) = 0
	\text.
\end{equation}
Again in order for these relations to be usable in the context of lattice Monte-Carlo we must integrate the fermions out. This time we find that these relations become trivial: the functions that have been shown to be $0$ have no dependence on $A$, and therefore cannot be used to remove the noise\footnote{An instructive example: beginning with \[
	0 = \int d\bar\psi d\psi \, \frac{\partial}{\partial \psi_i} \left[\psi_j e^{-\bar\psi D \psi}\right]\text,
\] differentiating and then taking the expectation value, we derive only that $D^{-1}_{ab} D_{bc} = \delta_{ab}$ (under the assumption $\det D \ne 0$).}.

Our final option is to derive Schwinger-Dyson relations before the auxiliary field is introduced; that is, while the action still contains a four-fermi term. We will not consider this procedure here for two reasons. First, because the introduction of the auxiliary field in the lattice formulation of the Thirring model used in this work in non-rigorous, meaning that Schwinger-Dyson relations derived before this step would, after the introduction of the auxiliary field, no longer hold exactly\footnote{In a similar vein one might consider deriving Schwinger-Dyson relations in the continuum and then using them on the lattice, counting on the fact that the bias introduced would vanish in the continuum limit. This approach is fraught: one might verify that the basis of control variates scales as $O(a)$, only to find that the optimized coefficients grow with $a^{-1}$ and the correct continuum limit is not reached.}. And second, because the fermionic theories of greatest interest in this work have gauge fields on the lattice, rather than mere auxiliary fields.

The usefulness of these Schwinger-Dyson relations, in the present context, is that they may be used to construct \emph{control variates}. If we are interested in the expectation value of an observable $\mathcal O$, we may subtract from it any function without changing the expectation value
\begin{equation}
	\langle \mathcal O - f \rangle = \langle \mathcal O \rangle
	\text,
\end{equation}
so long as we have an independent proof that $\langle f \rangle = 0$. While the expectation value is conserved by this subtraction, the noise is not, as can be seen from the variance:
\begin{equation}
	\langle (\mathcal O - f - \bar{\mathcal O})^2\rangle
	= \langle (\mathcal O - \bar{\mathcal O})^2\rangle
	- 2 \langle f (\mathcal O - \bar{\mathcal O})\rangle
	+ \langle f^2\rangle
	\text.
\end{equation}
The subtraction is most effective where the fluctuations in $f$ are strongly correlated with (and of the same magnitude as) those of $\mathcal O$. In this sense $f$ may be viewed as a model of the noise in $\mathcal O$. Furthermore, when $f$ is presented in the form of a total derivative (as in the construction of Schwinger-Dyson relations above), it is simultaneously a model of the noise and a ``certificate'' of the fact that it is, in fact, a model \emph{only} of noise.

It is trivial to see that an exact control variate---one that entirely removes statistical noise---always exists. For observable $\mathcal O$, this control variate is uniquely given by
\begin{equation}
	f_{\mathcal O} = \mathcal O - \langle \mathcal O \rangle
	\text.
\end{equation}
Less obvious is that such a control variate may always be obtained as a linear combination of total derivatives; that is, from Schwinger-Dyson relations. This is a direct consequence of the fact that, labelling the $n$-dimensional space of lattice configurations $M$, the top cohomology group $H^n(M;\mathbb R)$ is one-dimensional. If $M$ were disconnected with $k$ components, then we would have $H^n(M;\mathbb R) \approx \mathbb R^k$, with the $k$ dimensions corresponding to the $k$ integrals over the $k$ connected components. (A standard reference on this viewpoint is~\cite{Hatcher}.)

\subsection{Translation invariance}
Before proceeding further we must make our basis of control variates more concrete by selecting specific functions $g(A)$.  The simplest choice is to take $g(A) = 1$ to be constant. However, the resulting control variates are odd in $A$, and therefore vanish trivially by the $\mathbb Z_2$ symmetry $A \leftrightarrow -A$. By itself this does not imply that these control variates cannot be useful in reducing the noise of an observable $\mathcal O$. However, any observable of interest will be $\mathbb Z_2$-even (as the odd part exactly vanishes), and the $\mathbb Z_2$-odd control variate will have no overlap with the noise in an even function.

The next-simplest option is to take $g(A)$ to be linear in $A$ at small $A$. This results in a basis of control variates of dimension $4 V^2$:
\begin{equation}\label{eq:cv-v2}
	F_{\mu\nu}(x,x') e^{-S} \equiv \frac{\partial}{\partial A_\mu(x)} \left(\sin A_\nu(x')e^{-S}\right)
	\text.
\end{equation}
However, the observables of interest are all translationally invariant\footnote{Note in particular that a correlator is translationally invariant, being averaged across the entire lattice as an easy variance-reduction method.}. The noise we seek to model is therefore also translationally invariant, and we should restrict the space of control variates to the $4 V$-dimensional translationally invariant subspace of that defined by Eq.~(\ref{eq:cv-v2}). These are defined by
\begin{equation}\label{eq:cv-basis}
	F_{\mu\nu}(x) e^{-S} \equiv \sum_y \frac{\partial}{\partial A_\mu(y)} \left(\sin A_\nu(x+y)e^{-S}\right)
	\text.
\end{equation}
For small field values $A$ (so that $\sin A \sim A$), these control variates are quadratic. In fact this set of control variates provides a complete basis: any orthogonal control variate must be higher-order in $A$. Consequently this provides a complete model of the noise that appears at order $g^2$---this will be elaborated on in Sec~\ref{sec:perturbative}.

\subsection{Optimization}\label{sec:optimization}
Although as mentioned an exact control variate (one that entirely removes fluctuations in the observable) is guaranteed to exist, in practice we will not attempt to construct an exact control variate, but only an optimized one that is good enough for the task at hand. The construction of a quadratically complete basis in Eq.~(\ref{eq:cv-basis}) is a good starting point. We will find the optimal control variate in the span of that basis.

\begin{figure}
	\centering
	\includegraphics[width=0.95\linewidth]{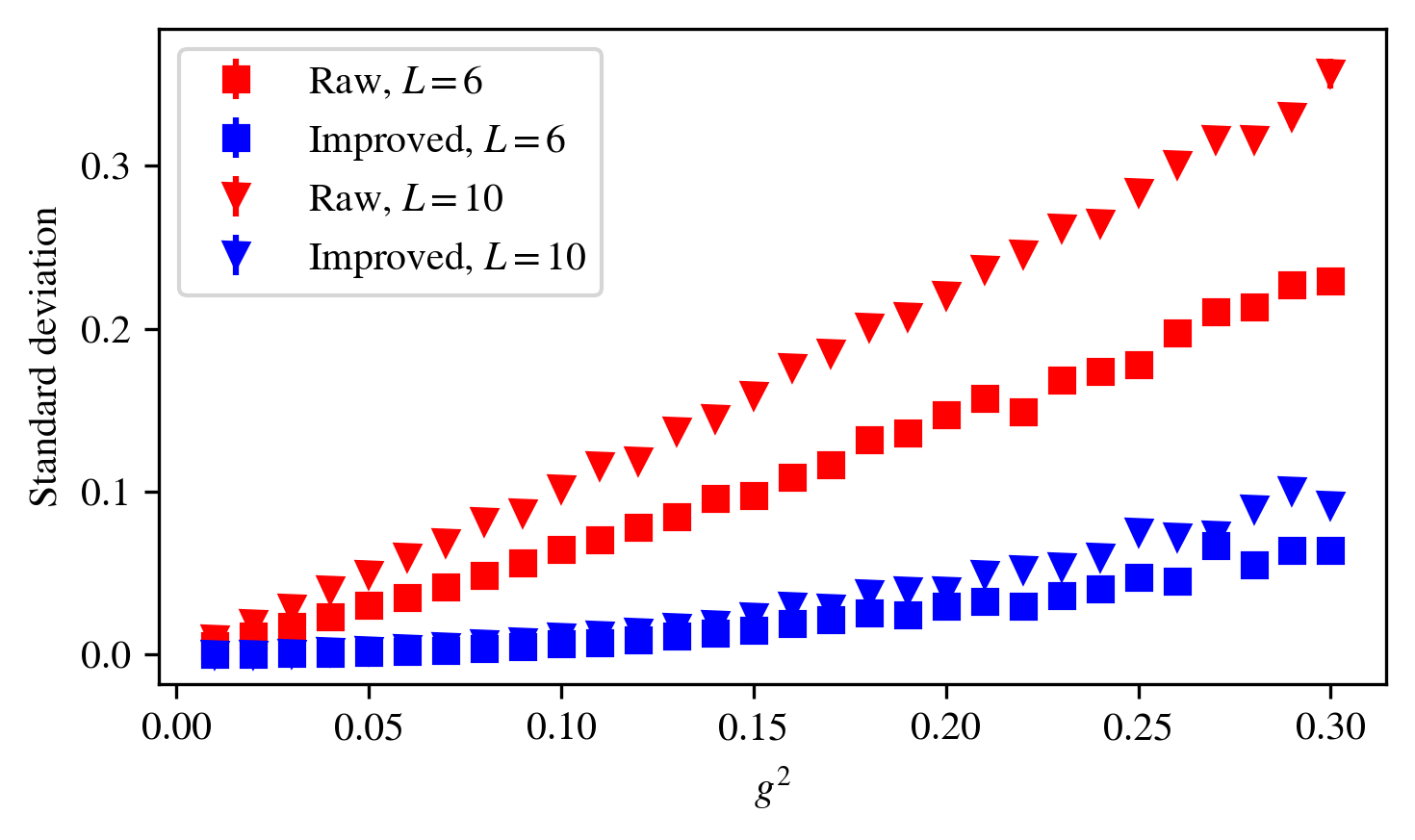}
	\caption{Standard deviation of the naive and improved estimators as a function of the coupling. In the Thirring model, the noise of both estimators vanishes in the weak coupling limit; the noise of the improved estimator goes as $g^4$, an improvement over the raw estimator scaling of $g^2$. Each data point uses $2 \times 10^3$ samples; in the improved estimator, half of those samples are used for fitting the control variate.\label{fig:coupling}}
\end{figure}

The optimization procedure was described in~\cite{Bhattacharya:2023pxx}---we repeat it here for completeness, as it is the inspiration for the perturbative procedure of the next section. Writing the basis of control variates with a single index---$F_i$---for brevity, consider the consequences of modeling the noise in $\mathcal O$ by the linear combination $f \equiv \sum_i c_i F_i$. The variance of this improved estimator is
\begin{equation}
	\mathrm{Var}[\mathcal O - f] = \langle (\mathcal O - \bar{\mathcal O})^2 \rangle - 2 c_i \langle \mathcal O F_i \rangle + c_i c_j \langle F_i F_j\rangle
	\text.
\end{equation}
This is a quadratic function in the coefficients $c$. It is convenient to package the correlations between different basis variates $F$ in a matrix $M$, and those between the basis variates and the observable in a vector $b$:
\begin{equation}
	M_{ij} = \langle F_i F_j\rangle \text{ and } b_i = \langle F_i \mathcal O \rangle
	\text.
\end{equation}
With this notation, where $M$ is non-singular the variance is minimized with the choice
\begin{equation}
	c = M^{-1} b
	\text.
\end{equation}
In practice we do not have exact expressions for either $M$ or $b$. They can of course be estimated from a Monte Carlo calculation. This then provides a Monte Carlo estimate for the approximately optimal coefficients $c$ defining the control variate, and this control variate is used to improve the estimation of the observable.

Two technical concerns arise when following this procedure. First, the matrix $M$ (either exact or estimated) is often ill-conditioned\footnote{However even the Monte Carlo estimate is guaranteed to be positive semi-definite.}. It is therefore typically necessary to add some regularization. An easy choice is $L_2$ regularization; in this work we accomplish this by replacing $M \rightarrow M + 10^{-2} I$. In~\cite{Bhattacharya:2023pxx}, $L_1$ regularization was used to cause $c$ to be sparse and avoid overfitting. In Section~\ref{sec:perturbative} we will present a different method towards the same goal: we perform perturbative calculations to remove the need for $\sim V$ fitting parameters.

Second, we must ensure that the estimation of $M$ and $b$ is uncorrelated with the subsequent estimation of $c_i F_i$ used to suppress the noise in $\mathcal O$. In practice we accomplish this by allocating some fraction (in this work, one-half) of the samples collected to the task of determining the optimal control variate, and only using the remainder for estimating $\mathcal O$.

An observable of central interest is the correlator $\langle \bar\psi(x)\psi(0)\rangle$, whose exponential decay reveals the spectrum of fermionic excitations. We define a zero-momentum, translation-invariant projection of this correlator as
\begin{equation}\label{eq:correlator}
	C(\tau) = \sum_{\substack{t,t'\\x,x'}} \delta_{\tau,(t-t')} (-1)^{\theta(t' - t) + (x+x')t} D^{-1}_{(0,x)(\tau,x')}
	\text.
\end{equation}
Figure~\ref{fig:coupling} shows the noise in the estimation of $C(0)$ as a function of the coupling $g^2$, at bare mass $m=0.2$, on both $6\times 6$ and $10 \times 10$ lattices. It is plain that the noise with the naive estimator scales as $g^2$, whereas the improved estimator has higher-order noise (in fact it is of order $g^4$). This is a consequence of the fact, previously mentioned, that our basis of control variates is complete in the set of quadratic functions.

\section{Perturbative control variates}\label{sec:perturbative}
The optimization method described above is effective, but impractical on large lattices. A typical QCD ensemble might have $\sim 10^3$ configurations available, with $\sim 10^8$ degrees of freedom in a basis analogous to that of Eq.~(\ref{eq:cv-basis}). By a wide margin, there is not enough information to determine the optimal coefficients defining the control variate.

Happily, other sources of information about the lattice theory are available beyond the configurations themselves. In this section we will see how a high-quality control variate can be determined from perturbation theory (a notion suggested in~\cite{Lawrence:2020kyw} with respect to the sign problem), with no fitting performed and no input from the Monte Carlo. Critically, the construction of the basis elements remains exactly as it was. Their nonperturbative construction guarantees that any linear combination has expectation exactly zero. Perturbation theory is used only to estimate the optimal coefficients. Where perturbation theory is a poor approximation, we will obtain a poor control variate, but no bias will be introduced into the results.

\begin{figure}
	\centering
	\includegraphics[width=0.95\linewidth]{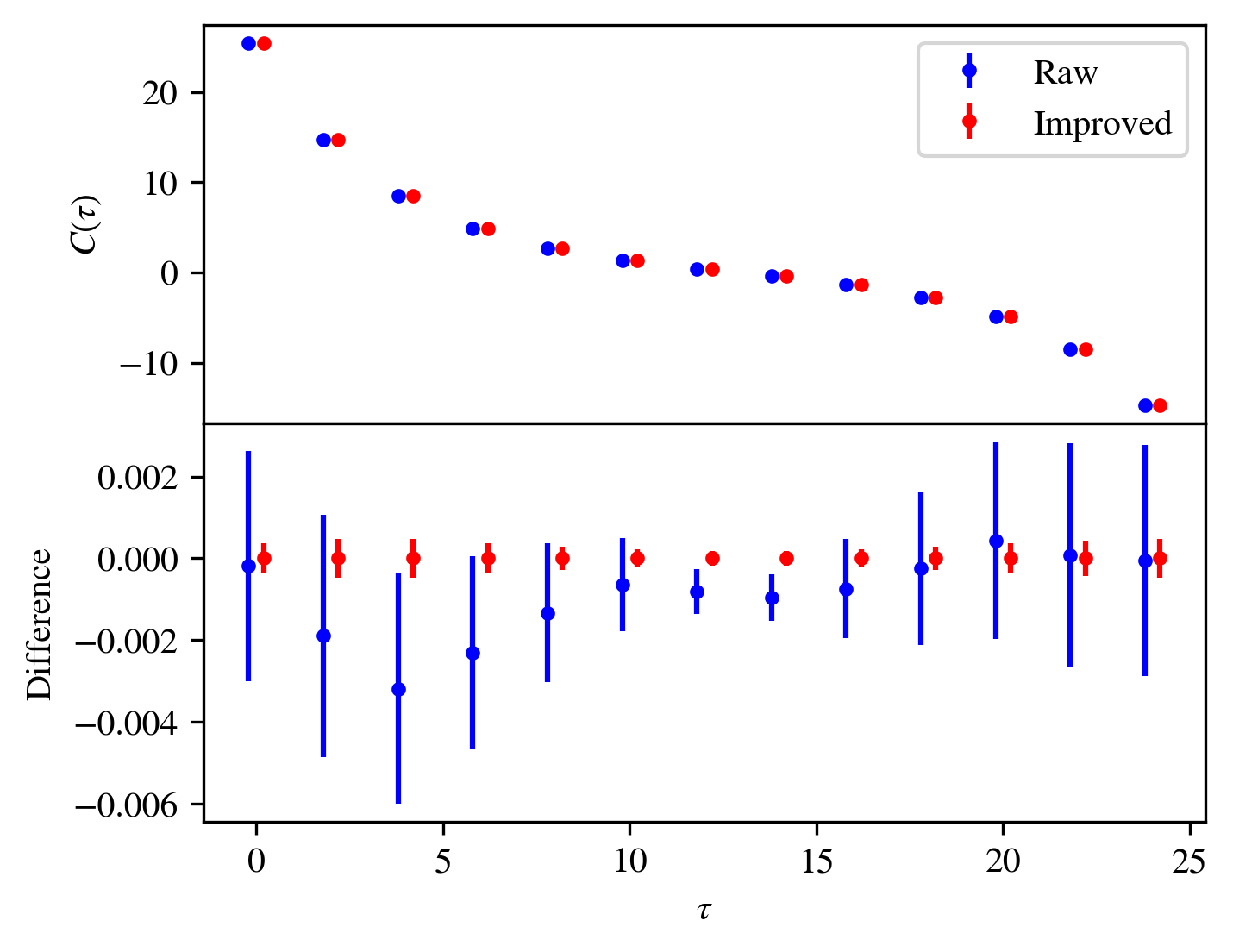}
	\caption{Fermion correlation function of the weak-coupling model, on a $26 \times 26$ lattice, computed with the naive and perturbative estimators. The bottom panel shows the difference between the raw and improved estimators, normalized so that the central value of the improved estimator is always at $0$. This calculation includes $5 \times 10^2$ samples.\label{fig:perturbative}}
\end{figure}
As before, let $\mathcal O$ be the observable of interest. We wish to find a function $g(\cdot)$ of the fields such that the noise of the subtracted observable $\tilde{\mathcal O} \equiv \mathcal O - \partial g + g \partial S$ is minimized. Several approaches present themselves. One is to recall from above that for every observable there is an exact control variate, and moreover that such a control variate may always be written in the form Eq.~(\ref{eq:cv-general}). This provides us with a differential equation, which schematically reads
\begin{equation}
	\mathcal O - \langle \mathcal O \rangle = \partial g\text.
\end{equation}
As a standard technique, this equation may be solved order-by-order in perturbation theory to reveal an optimal control variate.

Another possibility is to express the noise as a functional of $g$, and
minimize by demanding that the functional derivative with respect to $g$
vanish. It may be seen that this is equivalent to the procedure above.

A third approach is more straightforward, analogous to the nonperturbative optimization procedure followed above, and more likely to generalize to larger lattices. We choose in advance a basis from which our control variate will be constructed, compute perturbatively the correlation matrices $M$ and $b$ defined above, and obtain the coefficients $c$ as $c = M^{-1} b$ (with the inversion regularized as necessary).

When using the basis of Eq.~(\ref{eq:cv-basis}), it happens that this procedure will remove all noise at leading order in perturbation theory. Critically, this is not necessary in order for this method to be effective. The perturbative estimate of the optimal coefficients is useful even if the resulting control variate does not model all noise at a given order. For larger lattices, a basis that removes all noise at a given order in $g^2$ may be prohibitively expensive to compute. In this approach, it is instead sensible to write down the largest basis which is (relatively) cheap to compute, and use perturbation theory to find the best control variate available in that vector space.

Note also that the role of perturbation theory here is relegated to the computation of the correlations $M$ and $b$. Another approximation can be substituted without changing the rest of the procedure.

We now proceed to follow this procedure for perturbation theory in the Thirring model. Perturbation theory begins with the substitution $A = g B$, to convert the action to a Gaussian action plus higher-order corrections. (This substitution makes obvious why control variates that were low-order polynomials in $A$ removed all noise at low orders in $g$.)

Computing the correlation matrix $M$ and vector $b$ require expanding $S$, $F_{\mu\nu}(x)$, and $D^{-1}$ in $g^2$. We will obtain the optimal coefficients $c$ at order $g^2$ (which is the leading order).

We begin by keeping only the leading-order term in the action---higher-order terms do not contribute to $c$ at order $g^2$:
\begin{equation}
	S = \sum_x B_{\mu}(x)^2 + O(g)
	\text.
\end{equation}
Expectation values with respect to this action are therefore given by the rule
\begin{equation}
	\langle B_\mu(x) B_\nu(y)\rangle = \frac 1 2 \delta_{\mu\nu} \delta_{xy}
\end{equation}
combined with Wick's theorem.

We must also expand the control variates and the observable, this time both through order $g^2$. The control variates are found to be
\begin{equation}\label{eq:Fmunu}
	F_{\mu\nu}(x) = V \delta_{\mu\nu} \delta_{x0} - 2 \sum_y B_\mu(x+y) B_\nu(y)\text,
\end{equation}
where $V = L^2$ is the spacetime volume of the lattice. The inverse of the Dirac matrix is given by a standard formula:
\begin{widetext}
	\begin{equation}\label{eq:Dinv}
	D^{-1} = 
	D^{-1}_0 - g \Tr D^{-1} \frac{\partial}{\partial g}D
	+ g^2\left(\Tr D^{-1}_0 \frac{\partial D}{\partial g} D^{-1}_0 \frac{\partial D}{\partial g} D^{-1}_0 - \frac 1 2 \Tr D^{-1}_0 \frac{\partial^2 D}{\partial g^2}\right) + O(g^3)
	\text.
\end{equation}
\end{widetext}
Above, $D^{(0)}$ denotes the free Dirac matrix; that is, the Dirac matrix in a constant gauge background $A=0$. The term of order $g$, being both odd in the gauge field and imaginary, will not contribute.

	The above expression is not complete without the expansion of $D$ in powers of $g$. Both the first and second derivatives are readily evaluated from Eq.~(\ref{eq:dirac}) by making the same $A = g B$ substitution.

At this point it is useful to note that because the observable ($D^{-1}$) is constant at order $g^0$, the correlation vector $b \sim \langle F \mathcal O \rangle$ vanishes at that order. Therefore only the $g^2$ term in $b$ will appear in this calculation, and since we want the $g^2$ term of $c = M^{-1} b$, we need only compute the correlation matrix $M$ at order $g^0$. The result is:
\begin{equation}
	M_{\mu\nu}^{\mu'\nu'}\!(x',x) = V \left[
		\delta_{\mu\mu'}\delta_{\nu\nu'}\delta_{xx'} + \delta_{\mu\nu'} \delta_{\mu'\nu} \delta_{x',-x}
		\right]
	\text.
\end{equation}
To view this as a matrix (instead of a $4$-tensor-valued function of the Cartesian product of the lattice with itself), we group $(\mu,\nu,x)$ as one index and $(\mu',\nu',x')$ as the other.

Computing the vector $b_{\mu\nu}(x;z,z') \equiv \langle F_{\mu\nu}(x)D^{-1}_{z,z'}\rangle$ requires summing over all contractions between Eq.~(\ref{eq:Fmunu}) and Eq.~(\ref{eq:Dinv}). This may be accomplished programmatically; there are $\sim L^2$ terms in each of $\sim L^2$ components. The correlation function being examined, defined in Eq.~(\ref{eq:correlator}), imposes an additional factor of $L^3$. The total cost of this computation therefore scales as $L^7$, and this is the dominant cost in determining the coefficients defining the perturbative control variate.

Figure~\ref{fig:perturbative} shows the correlation function for the weak-coupling model defined in Section~\ref{sec:thirring} above. The raw and perturbatively improved estimators are in good agreement, with error bars made an order of magnitude smaller by the perturbative improvement. Note that, with a dense fit as described in the previous section, no fewer than $2\times 26 \times 26 \gtrsim 10^3$ samples would be needed to avoid overfitting. Only $5 \times 10^2$ samples are used in this calculation.

\section{Few-parameter fits}\label{sec:fitting}
We have so far detailed two procedures for determining the coefficients of $F_{\mu\nu}(x)$ defining a control variate. In Section~(\ref{sec:optimization}) we performed a statistics-intensive fit of all parameters; this procedure is impractical on large lattices, where the number of samples is dwarfed by the number of degrees of freedom in the fit. In the section above, we computed the coefficients at leading order in perturbation theory; one cannot expect this procedure to perform well in a nonperturbative regime (as lattice QCD is).

\begin{figure}
	\centering
	\includegraphics[width=0.95\linewidth]{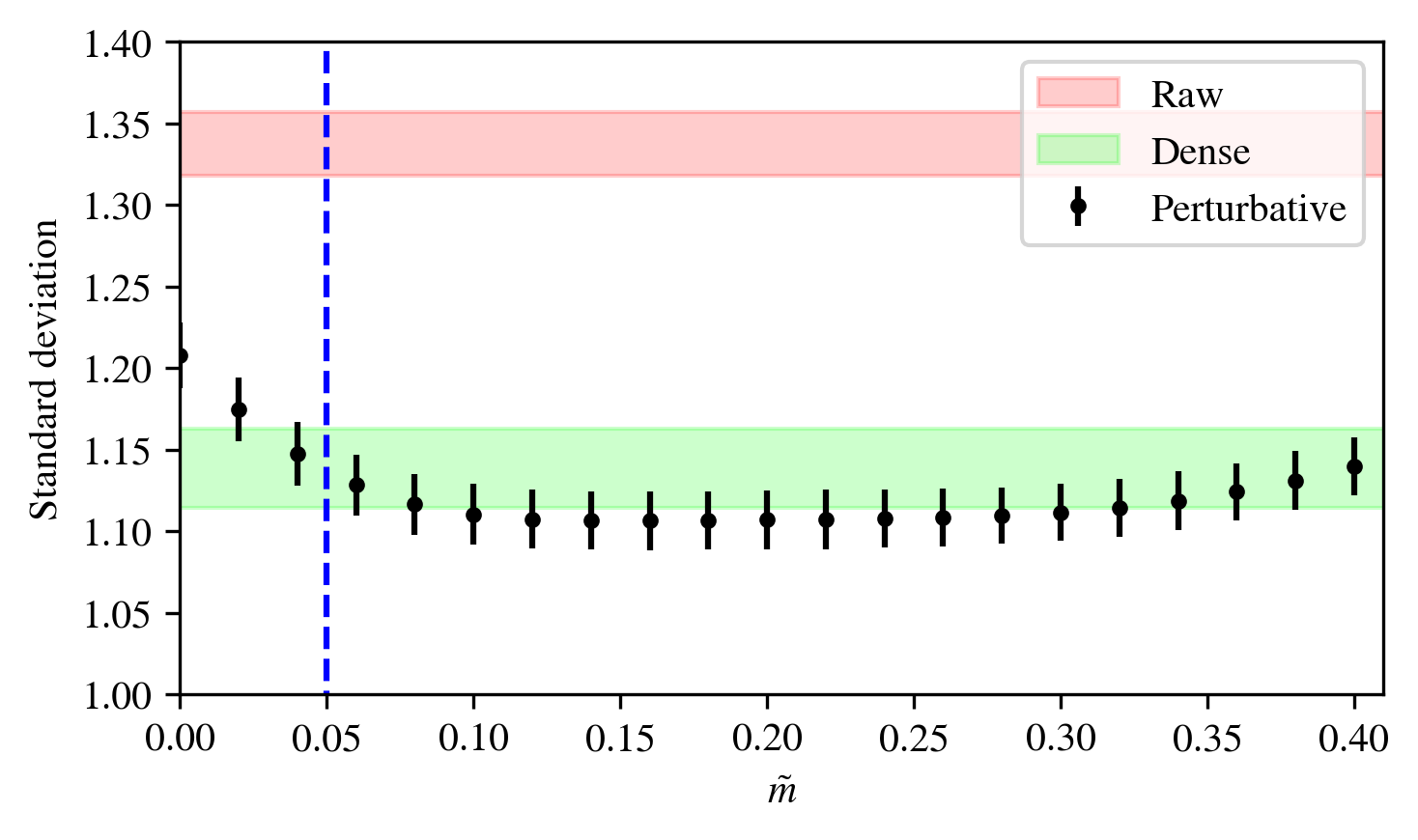}
	\caption{Standard deviation of the improved estimator of Section~\ref{sec:fitting} as a function of the parameter $\tilde m$ in the control variate. (Optimization over values of $\tilde g^2$ is performed separately for each data point.) The perturbative value is marked by the dashed blue line, and is clearly not the optimum. The red shaded region indicates the standard deviation obtained by the unimproved estimator; the green shaded region indicates that obtained by an estimator optimized according to the procedure described in Section~\ref{sec:optimization}.\label{fig:fit}}
\end{figure}

It is natural to attempt to use the perturbative calculation to inform the construction of a low-dimensional family of control variates, within which we can optimize without requiring $\sim V$ samples. This section is a first step towards that goal. We will not modify the basis of control variates $F_{\mu\nu}(x)$ defined by Eq.~(\ref{eq:cv-basis}), but introduce two free parameters into the calculation of their coefficients $c_{\mu\nu}^{(x)}$.

The perturbative calculation of $c_{\mu\nu}^{(x)}$ implicitly involves the lattice parameters defining the action: $g^2$ and $m$. In our leading-order calculation the coupling $g^2$ appears only as an overall coefficient; $m$ appears in a less trivial, and nonlinear, manner. Based on the calculation described in the previous section we may define coefficients $c_{\mu\nu}^{(x)}(\tilde g^2,\tilde m)$ obtained by substituting $\tilde g^2$ for $g^2$ (as the overall coefficient), and $\tilde m$ for $m$ (in $D_0^{-1}$ in Eq.~(\ref{eq:Dinv})). This defines a control variate with two natural free parameters to optimize.

Figure~\ref{fig:fit} shows the standard deviation in the estimate of $C(\tau = 5)$ on an $8 \times 8$ lattice using the strong-coupling parameters defined in Section~\ref{sec:thirring}. The standard deviation is shown as a function of only one parameter, $\tilde m$; for each value of $\tilde m$ the optimization over $\tilde g^2$ has already been performed. The standard deviation of the unimproved estimator of $C(\tau = 5)$, as well as that obtained via the procedure of Section~\ref{sec:optimization} are also shown.

Several surprising features of this procedure are apparent. The optimal value of $\tilde m$ is not the perturbative value (marked by the dashed line), but somewhat above it. At the same time, the standard deviation is remarkably insensitive to the value of $\tilde m$ across a wide range. Finally, this two-parameter fit, based on a perturbative calculation, obtains as much of an improvement at strong coupling as the fully nonperturbative optimization procedure used previously.

The range of optimal values of $\tilde m$ includes the renormalized mass, measured to be $\sim 0.25$. This is mildly suggestive, and reminiscent of an observation made regarding contour deformation methods that the optimal contour isn't the Lefschetz thimble given by the perturbative critical point, but one incorporating quantum corrections to that surface~\cite{Gantgen:2023byf}.

\begin{figure}
	\centering
	\includegraphics[width=0.95\linewidth]{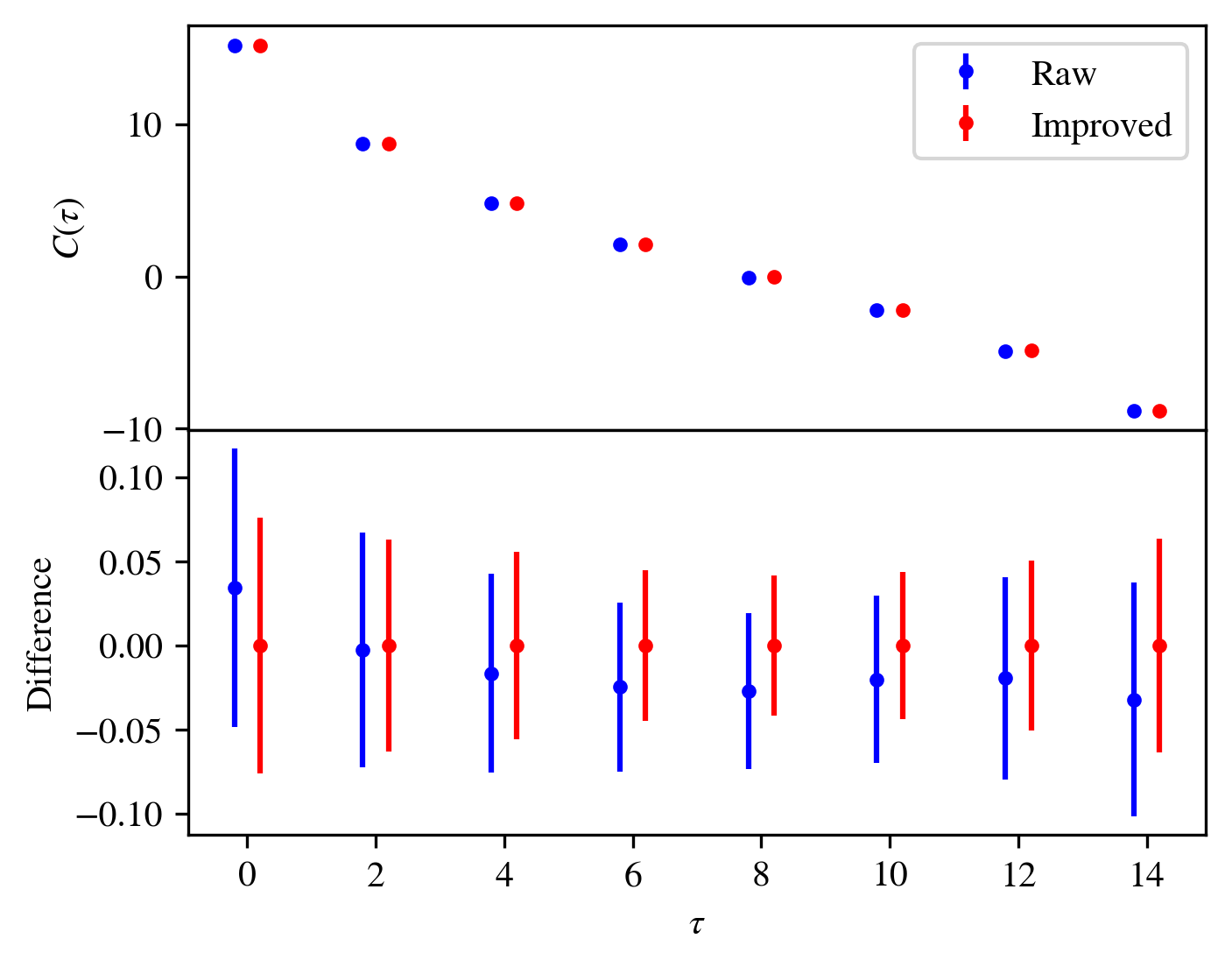}
	\caption{Fermion correlation function of the strong-coupling model, on a $16 \times 16$ lattice, computed with the naive estimator and a two-parameter fit improved estimator as described in the text. The bottom panel shows the difference between the two estimators, normalized so that the improved error bars are always centered at zero. A total of $5 \times 10^2$ samples are used to generate this plot.\label{fig:strong}}
\end{figure}

Unfortunately at strong coupling neither this procedure nor the full optimization procedure, when working with the basis of Eq.~(\ref{eq:cv-basis}), are able to achieve a substantial improvement in the signal-to-noise ratio. Figure~\ref{fig:strong} shows the full correlator on a $16 \times 16$ lattice using both the raw estimator and an improved estimator with $\tilde m = 0.2$ chosen arbitrarily. The error bars are consistently about 10\% smaller when the improved estimator is used.

\section{Discussion}\label{sec:discussion}
We have extended the method of~\cite{Bhattacharya:2023pxx} to theories with lattice fermions. Furthermore we have shown how this method of constructing control variates may be practical in a sample-constrained context, using perturbation theory to motivate functional forms for high-quality control variates.

Nothing in this work or its predecessors~\cite{Bhattacharya:2023pxx,Bedaque:2023ovz} achieves even a factor-of-$2$ reduction in the signal-to-noise problem of a strongly coupled theory without a many-parameter fit. In Section~\ref{sec:perturbative} we saw that a perturbative calculation of the optimal control variate was highly effective at weak-coupling. Meanwhile, a central accomplishment of~\cite{Bhattacharya:2023pxx} was the dramatic increase of the signal-to-noise ratio even at strong coupling by a many-parameter fit\footnote{Note that the sparse optimization performed in that work was demonstrated only at relatively weak coupling, and moreover that there were still $\sim 10^2$ parameters used on a two-dimensional lattice. Assuming even mild scaling with lattice volume, this will not be possible in the context of lattice QCD.}.

This paper, like previous work in this line (e.g.~\cite{Bhattacharya:2023pxx,Bedaque:2023ovz}), has not addressed the exponential scaling of the signal-to-noise problem. The reason for this is best understood by comparison with sign problems. Signal-to-noise problems and sign problems are closely related. Section~\ref{sec:introduction} touched on a separation between the two in their asymptotic scaling: typical sign problems scale with spacetime volume (or number of lattice sites), while many signal-to-noise problems have difficulty proportional only to a length (often separation in a correlator). As a result, effective techniques for dealing with the two problems differ wildly. No modern computation comes close to being able to estimate, directly from lattice QCD, the zero-temperature equation of state at high densities. Any effective attack on this problem must proceed by substantially modifying the exponential scaling, and conversely any algorithm that does modify the exponential scaling will outperform all known algorithms. Signal-to-noise problems, by contrast, are often mild, with desired physical results ``inaccessible'' by only a factor of $\lesssim 10$. A collection of computationally cheap improvements, each yielding a factor of $\sim 2$ in effective statistics, can easily accomplish this without the asymptotic scaling of the problem being affected.

\begin{acknowledgments}
	I am grateful to Tanmoy Bhattacharya, Rajan Gupta, and Jun-Sik Yoo for several useful conversations, and their encouragement over the course of this project.

This work was supported by a Richard P.~Feynman fellowship from the LANL LDRD program. Los Alamos National Laboratory is operated by Triad National Security, LLC, for the National Nuclear Security Administration of U.S. Department of Energy (Contract No. 89233218CNA000001).
\end{acknowledgments}

\bibliography{refs}

\end{document}